# Ligand Dependent Oxidation Dictates the Performance Evolution of High Efficiency PbS Quantum Dot Solar Cells

*David Becker-Koch[a], Miguel Albaladejo Siguan[a], Vincent Lami[a], Fabian Paulus[a], Hengyang Xiang[b], Zhuoying Chen[b] and Yana Vaynzof [a]\**

[a] Integrated Centre for Applied Physics and Photonic Materials and Centre for Advancing Electronics Dresden (cfaed), Technical University of Dresden, Nöthnitzer Straße 61, 01187 Dresden, Germany

*Corresponding author, e-mail: yana.vaynzof@tu-dresden.de

[b] LPEM, ESPCI Paris, PSL Research University, Sorbonne Université, CNRS, 10 Rue Vauquelin, 75005 Paris, France

**Abstract**

Lead sulfide (PbS) quantum dot (QD) photovoltaics have reached impressive efficiencies of 12%, making them particularly promising for future applications. Like many other types of emerging photovoltaic devices, their environmental instability remains the Achilles heel of this technology. In this work, we demonstrate that the degradation processes in PbS QDs which are exposed to oxygenated environments are tightly related to the choice of ligands, rather than their intrinsic properties. In particular, we demonstrate that while 1,2-ethanedithiol (EDT) ligands result in significant oxidation of PbS, lead iodide/lead bromide ($PbX_2$) coated PbS QDs show no signs of oxidation or degradation. Consequently, since the former is ubiquitously used as a hole extraction layer in QD solar cells, it is predominantly responsible for the device performance evolution. The oxidation of EDT-PbS QDs results in a significantly reduced effective QD size, which triggers two competing processes: improved energetic alignment that enhances electron blocking, but reduced charge transport through the layer. At early times, the former process dominates, resulting in the commonly reported, but so far not fully explained initial increase in performance, while the latter governs the onset of degradation and deterioration of the photovoltaic performance. Our work highlights that the stability of PbS quantum dot solar cells can be significantly enhanced by an appropriate choice of ligands for all device components.

## 1. Introduction

Since their first introduction in 2008,[1] lead sulfide (PbS) quantum dot (QD) photovoltaic devices have been under extensive investigation, leading to a record power conversion efficiency of 12.24%.[2–9] One of the key advantages of quantum dots stems from their nano-character, which allows for tuning of their optoelectronic properties by choosing their size, shape and ligands.[10–14] This multi-functionality has led to the realization that QDs can be used both as an active layer and as an extraction layer, provided that appropriate ligands are selected.[15] In particular, PbS QDs with 1,2-ethanedithiol (EDT) ligands have become ubiquitously employed as a hole extraction layer in PbS quantum dot photovoltaic devices, including those with record efficiency.[8,9]

Another unique trait of quantum dot solar cells is that, unlike other types of solution-processed devices, e.g. organic or organic-inorganic perovskite,[16–19] quantum dot solar cells often experience an increase in performance upon exposure to the environment. Consequently, it has become routine in literature to perform preconditioning procedures, such as exposure to air prior to photovoltaic characterization.[8,15,20–23] However, these procedures are not standardized and a clear understanding of the evolution of the photovoltaic performance upon exposure to environmental factors has not emerged.

As the main focus of the academic community has been on improving the device efficiency,[8,9,15,24–29] minor effort has been devoted to the study of their stability.[8,15,20–23] While many studies report that the photovoltaic characterization of PbS solar cells is carried out under inert conditions after exposure to air, it has been shown that prolonged exposure to $N_2$ strongly degrades the device performance.[30] Early reports investigating the interaction of PbS quantum dots with oxygen have shown that the surface of PbS QD may oxidize, leading to the formation of lead oxide (PbO), lead sulfite ($PbSO_3$), and lead sulfate ($PbSO_4$), where the ratio of the latter two degradation products depends on the size of the quantum dots.[31,32] This oxidation process has been linked to the formation of a thin outer shell which effectively reduces the size of the dot,[33] and is associated with a reduction in the leakage current and the suppression of bimolecular recombination.[31] Early reports also investigated the evolution of trap states upon oxidation, showing that trap states are either filled or passivated by oxidation.[34,35] Studies on the effect of humidity are even rarer, with a recent report by Kuwarmi *et al*. showing that the humidity present during fabrication of the devices strongly affects their performance.[36] Other reports often mention shelf-storage stability, with impressive demonstrations of up to half a year stable PCE of non-encapsulated devices,[8,15] but these experiments do not allow for a controlled environment and do not elucidate the mechanisms behind the observed performance evolution.[37] Most critically, while shelf-storage stability is an important indicator for the viability of PbS solar cells for industrial applications, stability under continuous illumination is an even more crucial aspect for future advancement of this technology. The lack of continuous light degradation studies for PbS QDs cells limits our understanding of degradation mechanisms in these devices and hinders the development of mitigation strategies.

Herein, we characterize the evolution of the photovoltaic performance of high efficiency PbS solar cells upon exposure to controlled environments (pure $N_2$, $N_2$ + 20% relative humidity (RH), $N_2$ + 20% $O_2$ and simulated air) under continuous illumination. We observe that in oxygenated environments the device performance evolves in three distinct phases: first, a sharp increase in all photovoltaic parameters that occurs on the time scale of minutes, followed by a second slower, more gradual performance increase, and concluded by a third phase in which the performance progressively deteriorates. These three phases are observed in both dry and humid oxygenated environments, with the addition of humidity prolonging the duration of the second phase. By employing a range of spectroscopic methods, we demonstrate that the presence of the second phase is associated with different rates of oxidation of the PbS quantum dots in the active and extraction layers of the device, with the latter being responsible for both the improved performance in phase II and the loss of performance in phase III. Remarkably, we demonstrate that the active layer itself is robust against degradation, demonstrating that it is the choice of ligands that determines the stability of PbS QDs. Our study highlights the need to replace the commonly used EDT-coated QD hole extraction layer in the device structure to enhance device stability.

2. **Experimental**

PbS QD synthesis followed Bakulin *et al.* [38] decreasing the injection temperature from 125°C to 90°C to achieve the range of different sizes, respectively.

Substrate, both glass and ITO, preparation followed Weu *et al.* [17], as well as the fabrication of the ZnO sol-gel layer. $PbX_2$-PbS and EDT-PbS layer deposition was adapted from Liu *et al.* [8], noting that the concentrations in use changed a bit depending on the size of the dots. The gold contacts where thermally evaporated, leading to a solar cell area of 4.5mm$^2$

Degradation was carried out in a self-built environmental rig, which is capable of being load from a $N_2$ glovebox where the samples were stored. The amount of nitrogen, oxygen and humidity in the rig was monitored during measurements. For nitrogen only flows, the base content of $O_2$ and $H_2O$ was in the low ppm while for degradation runs, 20-25% $O_2$ and/or 20% RH was set. The artificial AM1.5 sunlight at 100 mW/cm$^2$ (no mask, no mismatch factor (MF) correction, normally MF~1.05) was provided by an Abet Sunlight Class A solar simulator. For light intensity dependence measurements reflective neutral density filters were used to decrease the light intensity on the cells.

Photovoltaic parameters where recorded using a Keithley 2450 source measure unit while the samples were in the environmental rig and under the solar simulator. Both forward and backward bias direction were recorded. The number of measured cells changed, depending on the gain. The minimal number for evaluation was five for the small dot degradation and the maximum eleven in three batches for the different atmosphere study.

Absorptions measurements were done in air with a JASCO UV-vis-spectrometer V-770, but the probes were directly taking form their corresponding atmospheres, such that no transport time was in between. The NIR absorption peak position was extracted by fitting a Gaussian minus background. To estimate the dot radius the TEM picture in Figure S1 was used.

The QD films investigated for PES measurements were fabricated the same way as the corresponding photovoltaic devices. After the films were finished, they were stored in a nitrogen glovebox before being transferred into an ultrahigh vacuum (UHV) chamber of the PES system (Thermo Scientific ESCALAB 250Xi) for measurements. The samples were exposed to air only for a short time span of approximately 30 seconds. All measurements were performed in the dark. UPS measurements were carried out using a double-differentially pumped He discharge lamp ($h\nu = 21.22$ eV) with a pass energy of 2 eV and a bias at -5 V. XPS measurements were performed using an XR6 monochromated Al K$\alpha$ source ($h\nu = 1486.6$ eV) and a pass energy of 20 eV. The Transmission Electron Microscope (TEM) pictures were taken with a JEOL 2010 TEM. Suspended nano-particles were drop-cast on a TEM copper-grid. The grid has a thin supporting layer of a FormVar® film, topped with amorphous carbon. The characterization started only after complete solvent evaporation.

The X-ray Diffraction Spectroscopy (XRD) measurements of the quantum dot films were conducted with a Rigaku SmartLab diffractometer with a 9kW rotating copper anode. 2D intensity maps were recoded using a 2D HyPix3000 detector in a coupled theta-2theta scan (beam collimator 0.5mmφ). The map was background corrected and a central profile was taken to obtain the intensity vs. 2theta diffractogram, which was normalized to account for variations in film thickness. Contributions from K$\alpha$2 line were stripped using the SmartLab Studio II software.

### 3. Results and Discussion

### 3.1 Photovoltaic performance under different environmental conditions

In our study, we focus on high performance devices adapting the structure (Figure 1a) first reported by Liu *et al.*[8] In this structure, a thin film comprised of PbS QDs, capped by a mixture of $PbI_2$ and $PbBr_2$ ligands (termed $PbX_2$), is deposited in a single step as the device active layer, on top of which is a hole extraction layer of EDT coated QDs (Figure 1b). Figure 1c shows that this structure results in devices with high power conversion efficiencies –surpassing 10%– in agreement with previous reports.[8,9,36]

Figure 2 shows the evolution of the photovoltaic performance, normalized to the initial performance, directly after fabrication in various controlled environments and under continuous illumination. Exposure to an inert (meaning here an oxygen free, nitrogen only) atmosphere results in a sharp decrease in efficiency, which stabilizes at 25% of the original value after approximately 5h. The introduction of humidity does not strongly affect the dynamics of performance loss, with a similar final value of ~20% of the initial performance. However, oxygen exposure has a drastic effect on the device performance evolution. For both dry and humid oxygenated environments, a sharp increase in performance is observed almost instantaneously (phase I). This initial rise is followed by a gradual continuous increase on a time scale of hours, the exact duration of which is dependent on the presence of humidity (phase II). In both cases the efficiency improves by approximately 90%, reaching similar maximum values. Finally, the performance starts to deteriorate on the time scale of tens of hours (phase III).

Our results demonstrate the extreme sensitivity of the device performance to preconditioning. While it is common to expose the devices to air prior to measurement under inert conditions, many parameters of this procedure will affect the measured photovoltaic performance. For example, the duration of exposure to air, the presence of light and the relative level of humidity will all influence the device performance.

Moreover, performing the photovoltaic characterization under inert conditions might also alter the performance, as we observe a sharp decrease in efficiency upon exposure to $N_2$.

### 3.2 Effect of oxygen exposure

The initial boost in efficiency (phase I) has been attributed to the formation of a thin shell of oxidation products surrounding the quantum dots[31,36] and passivation of surface trap states.[34,35] This, amongst other processes, reduces the bimolecular recombination in the device.[31] We also observe a reduction of bimolecular recombination by measuring the light intensity dependency of the $V_{OC}$ [39], shown in Figure S2. Pristine devices show strong bimolecular recombination, which is reduced upon exposure to oxygen even for a short period of time (12 min). The increase in the photovoltaic performance observed in phase II, however, has not been reported in literature to date. In this work we focus on identifying the origin of this secondary increase and note that while the addition of humidity in our experiments marginally slows down the effect, it does not change the overall mechanism behind it.

To understand the changes in quantum dot size upon oxygen exposure, we monitored the position of the absorption peaks of $PbX_2$-PbS and EDT-PbS QDs (initial peak position at 1010 nm) over time (Figure 3a). The behavior of the dots varies drastically depending on the type of ligand. The $PbX_2$-PbS QDs show a very small blue shift of ~10 nm (corresponding to a minor decrease in size) upon the initial exposure to oxygen, in agreement with known observations from phase I.[31,36] After this initial change, the dots' size remains approximately constant for the remainder of the experiment. EDT-PbS QDs, on the other hand, show a continuous reduction in size, with the absorption peak eventually reaching ~800 nm after 22.5h, over 200 nm blue shifted from their initial first excitonic absorption peak position. The different decrease in quantum dot size is also supported by X-ray diffraction (XRD) measurements (see Figure S3). While for $PbX_2$-coated dots the diffractogram remains unchanged after 3h of degradation in dry oxygen atmosphere, the diffractogram of the EDT-coated dots shows a significant broadening of the reflections, attributed to the shrinking size of the crystalline material.

This stark difference in the dynamics of size evolution for each of the two ligands is invariant with the initial dot size. Figure 3b shows the progression of the absorption peak corresponding to the first excitonic transition for three differently sized quantum dots (950nm, 1010 nm and 1050 nm, original data shown in Figure S4 & Figure 3a). In all three cases, the EDT coated dots strongly decrease in size, while the $PbX_2$ coated dots remain unchanged. These results are independent of the layer thickness, suggesting that oxygen penetrates the entire QD layer and that the process is not limited by oxygen diffusion (Figure S4c). The peak position as measured from external quantum efficiency (EQE) spectra remains stable for the first 10h, as it is dominated by the absorption of the $PbX_2$-PbS QDs in the active layer (Figure 3b & S4a). For longer durations, the peak begins to shift slightly to lower wavelengths due to the broadening introduced by the strong shift of the EDT-PbS QDs.

To understand the chemical processes taking place at the surfaces of the QDs with both types of ligands, we monitored the evolution of their composition upon oxygen exposure with X-ray photoemission spectroscopy (XPS), shown in Figure 3c and 3d. Also displayed are the single fits to the different species of sulfur, in Figure S5. In the case of pristine EDT-PbS QDs, a single species of Pb (associated with PbS at 137.6 eV) and two species of S are observed (assigned to PbS and bound thiolate, at binding energies of 160.6 eV and 161.7 eV, respectively). A single species of C originating from the EDT ligand and the absence of oxygen confirms the high quality of the inert EDT-PbS layers.[30] However, once the dots are exposed to oxygen, their chemical composition changes drastically. The Pb *4f* spectrum becomes broader and shifts to higher binding energies, associated with the formation of a Pb-(R)$O_x$ shell, and is thickened upon prolonged exposure to oxygen.[40–42] The S *2p* spectra reveal the formation of three additional S species, namely unbound thiolate (163.3 eV), $PbSO_3$ (165.9 eV) and $PbSO_4$ (167.8 eV).[31] It is interesting to note that the overall amount of thiolate (bound and unbound) does not change, suggesting that no EDT is lost during degradation (Figure S6). This is also supported by the C *1s* signal that remains nearly unchanged throughout the experiment. The O*1s* spectra show a substantial increase in both the PbO and Pb(R)O$_x$ species upon degradation.

Unlike EDT-PbS QDs, PbX$_2$-PbS shows only minimal changes in composition upon degradation, in agreement with the optical characterization described above. No significant changes in the Pb *4f* or S *2p* can be observed, suggesting that the PbS dots remain predominately intact (Figure S6). The C *1s* and O *1s* do show a small increase upon oxygen exposure, but these signals remain substantially lower than those corresponding to EDT-PbS dots. These results confirm that PbX$_2$-PbS QDs are significantly more stable in the presence of oxygen, suggesting that the changes in photovoltaic performance are predominantly associated with the processes taking place in the EDT-PbS hole extraction layer.

The evolution of the photovoltaic performance in a humid oxygenated environment (Figure 2) shows a similar, albeit slightly slower, increase in performance during phase II. To compare the changes to the chemical composition of the EDT-PbS QDs between the dry and humid oxygenated atmospheres, we performed XPS measurements on films exposed to simulated air for different durations (Figure S7 & S8). We observe that the presence of humidity reduces the amounts of lead sulfite (PbSO$_3$) and lead sulfate (PbSO$_4$) species, which supports the theory that the formation of the oxidized shell occurs on slightly longer timescales (Figure S6). One possible explanation for this observation is that the formation of the oxidized products shell is slowed down by the presence of -OH groups on the surface of the dot. However, it is important to note that this effect results in only a minor stabilization of the QD, merely postponing the onset of degradation by a matter of hours.

The significant decrease in size of the EDT-PbS dots has two implications for the photovoltaic performance. On the one hand, a reduction in effective size would result in an increase of the optical bandgap and shift in the energy levels of the EDT-PbS layer. At the same time, the formation of an oxidized shell will negatively affect charge transport through the layer. It is the interplay between these two effects which determines the dynamics of phases II and III.

To understand how the change in the effective size of the EDT-PbS QD affects the energetic alignment during phase II, we performed ultra-violet photoemission spectroscopy measurements (UPS) on pristine and degraded QD films that had been exposed to oxygen for 2h, which corresponds to the peak efficiency achieved in this phase. Figure 4 shows the UPS spectra and the corresponding energy level diagrams constructed by combining these measurements with the bandgaps extracted from the optical characterization in Figure 3b. The results show that while no change in the energetics of $PbX_2$-PbS layer occurs (within the experimental resolution), the degradation of the EDT-PbS layer results in a significant enhancement in the electron-blocking at the active layer/hole extraction layer interface. Pristine devices exhibit a small barrier ($\Delta E_{LUMO}$) of only 0.15 eV, insufficient for the efficient blocking of electrons, which results in high recombination losses. On the other hand, in phase II the effective dot size of the EDT-PbS extraction layer is reduced until an optimum energetic alignment is achieved after 2h, with an electron blocking barrier of 0.5 eV. We note that the size reduction process progresses beyond this point, such that the barrier continues to increase (see Figure S9). However, the onset of phase III suggests that the benefits of any further improvements in energetic alignment are negated by the deterioration of charge transport properties of the EDT-PbS layer due to the build-up of oxidation products. This onset is triggered once the oxidation products shell is too thick to allow efficient hole extraction and transport to the anode. The reduced charge transport can be identified in dark I-V measurements during the degradation, which are depicted in Figure S10. Devices without an EDT-PbS layer stay more conductive than those with an EDT-PbS layer.

To summarize, exposure to oxygen leads to two competing processes: improved energetics and reduced charge transport, both originating from an effective size reduction of the EDT-PbS dots in the extraction layer. The optimal performance point is achieved at the cross-over between the dominance of each effect.

### 3.3 Model of performance evolution upon oxygen exposure

Figure 5 illustrates a model that summarizes the mechanisms responsible for the evolution of the photovoltaic performance in all three phases. Prior to oxygen exposure (pristine), leakage pathways and bimolecular recombination result in significant losses in device performance. In phase I, these leakage pathways and recombination are suppressed both in the active and the extraction layers, leading to a sharp increase in the photovoltaic performance. This occurs via, for example, a passivation of surface trap states as well as other effects already discussed in previous works. [31,33–35,43]

After this initial phase, no further changes occur to the $PbX_2$-PbS active layer, but the size of the EDT-PbS QDs continuously decreases. This decrease improves the energetic alignment at the interface to the active layer, improving electron blocking and resulting in further enhancement of the device performance (phase II). With further degradation, the transport through the EDT-PbS extraction layer is suppressed and the device performance decreases (phase III).

It is interesting to note that the presence of phase II is not universal and is dependent on the size of the PbS dots. For example, Figure S11 shows that the evolution of device photovoltaic performance is dramatically different when they are fabricated from QDs of different sizes (initial absorption peaks at 1050 and 950 nm). The performance evolution of devices based on the smaller dots shows only phase I, followed directly by phase III, suggesting that the transport through these initially smaller dots is hindered due to the formation of the oxidation shell almost immediately after they are exposed to oxygen. Consequently, such a device does not benefit from the improved energetic alignment, since both electrons and holes cannot be transported through the oxidized EDT-PbS extraction layer. This is further supported by the optical measurements in Figure 3b, which show a faster reduction in size for initially smaller dots. Since the change in the initial size – which is necessary to alter the magnitude of the performance increase that occurs during phase II – is small, the increase could covertly contribute to the performance evolution of many high efficiency devices employing QDs around this common size range.

4. Conclusion

To summarize, our results demonstrate that it is the choice of ligands that determines the stability of PbS QDs upon exposure to oxygenated environments. In devices employing PbS QDs as both active and hole transport layer, it brings about two competing effects: enhanced electron blocking caused by the improvement of the energetic alignment at the $PbX_2$-PbS/EDT-PbS interface, and loss of charge transport through the unstable EDT-PbS layer. We show that early on, the first effect dominates and the photovoltaic performance is significantly enhanced. In later times, the latter effect becomes more pronounced resulting in a severe loss in performance. This work highlights the urgent need to replace the currently ubiquitously used EDT-PbS layer, as we show that it is neither energetically optimal as an electron blocking layer, nor is it stable enough to maintain charge transport upon exposure to oxygenated environments under continuous illumination. The use of novel, more robust hole-extraction layers, combined with the excellent stability of the $PbX_2$-PbS active layer will significantly advance both the efficiency and durability of PbS-based solar cells, taking them one step closer to industrial application.

**Conflicts of interest**

The authors declared no competing interests.

**Acknowledgements**

We kindly thank Prof. U. H. F. Bunz for access to device fabrication facilities. We thank Dr. X. Xu (ESPCI Paris) for her kind assistance in TEM experiments.

The project has received funding from SFB 1249 (Project C04) as well as the 'PROCES' project (DFG: Grant VA 991/2-1, ANR: 17-CE05-0028-01). This project has also received funding from the European Research Council (ERC) under the European Union's Horizon 2020 research and innovation programme (ERC Grant Agreement n° 714067, ENERGYMAPS).

**References**

1. E.J.D. Klem, D.D. MacNeil, L. Levina, E.H. Sargent, *Adv Mater*, 2008, **20**, 3433–9, 10.1002/adma.200800326

2. M.A. Hines, G.D. Scholes, *Adv Mater*, 2003, **15**, 1844–9, 10.1002/adma.200305395

3. I. Gur, N. a Fromer, M.L. Geier, A.P. Alivisatos, *Science*, 2005, **310**, 462–6, 10.1126/science.1117908

4. E.A.L. Ellingson Randy J., Beard Matthew C., Johnson Justin C., Yu Pingrong, Micic Olga I., Nozik Arthur J., Shabaev Andrew, R.J. Ellingson, M.C. Beard, J.C. Johnson, P. Yu et al., *Nano Lett*, 2005, **5**, 865–71, 10.1021/nl0502672

5. J.M. Luther, M. Law, Q. Song, C.L. Perkins, M.C. Beard et al., *ACS Nano*, 2008, **2**, 271–80, 10.1021/nn7003348

6. G. Konstantatos, L. Levina, A. Fischer, E.H. Sargent, *Nano Lett*, 2008, **8**, 1446–50, 10.1021/nl080373e

7. E.J.D. Klem, H. Shukla, S. Hinds, D.D. MacNeil, L. Levina et al., *Appl Phys Lett*, 2008, **92**, 90–3, 10.1063/1.2917800

8. M. Liu, O. Voznyy, R. Sabatini, F.P. García De Arquer, R. Munir et al., *Nat Mater*, 2017, **16**, 258–63, 10.1038/nmat4800

9. J. Xu, O. Voznyy, M. Liu, A.R. Kirmani, G. Walters et al., *Nat Nanotechnol*, 2018, **13**, 456–62, 10.1038/s41565-018-0117-z

10. P.R. Brown, D. Kim, R.R. Lunt, N. Zhao, M.G. Bawendi et al., *ACS Nano*, 2014, **8**, 5863–72, 10.1021/nn500897c

11. Z. Liu, J. Yuan, S.A. Hawks, G. Shi, S.T. Lee et al., *Sol RRL*, 2017, **1**, 1600021, 10.1002/solr.201600021

12. C.R. Kagan, E. Lifshitz, E.H. Sargent, D. V. Talapin, *Science*, 2016, **353**, 10.1126/science.aac5523

13. D. Kim, D.H. Kim, J.H. Lee, J.C. Grossman, *Phys Rev Lett*, 2013, **110**, 1–5, 10.1103/PhysRevLett.110.196802

14. B. Ehrler, B.J. Walker, M.L. Böhm, M.W.B. Wilson, Y. Vaynzof et al., *Nat Commun*, 2012, **3**, 10.1038/ncomms2012

15. C.H.M. Chuang, P.R. Brown, V. Bulović, M.G. Bawendi, *Nat Mater*, 2014, **13**, 796–801, 10.1038/nmat3984

16. Q. Sun, P. Fassl, D. Becker-Koch, A. Bausch, B. Rivkin et al., *Adv Energy Mater*, 2017, **7**, 10.1002/aenm.201700977

17. A. Weu, J. Kress, F. Paulus, D. Becker-Koch, V. Lami et al., *ACS Appl Energy Mater*, 2019, acsaem.8b02049, 10.1021/acsaem.8b02049

18. H.J. Snaith, *Nat Mater*, 2018, **17**, 372–6, 10.1038/s41563-018-0071-z

19. W.R. Mateker, M.D. McGehee, *Adv Mater*, 2017, **29**, 10.1002/adma.201603940

20. Z. Ren, Z. Kuang, L. Zhang, J. Sun, X. Yi et al., *Sol RRL*, 2017, **1**, 1700176, 10.1002/solr.201700176

21. R. Azmi, S.Y. Nam, S. Sinaga, S.H. Oh, T.K. Ahn et al., *Nano Energy*, 2017, **39**, 355–62, 10.1016/j.nanoen.2017.07.015

22. S. Pradhan, A. Stavrinadis, S. Gupta, S. Christodoulou, G. Konstantatos, *ACS Energy Lett*, 2017, **2**, 1444–9, 10.1021/acsenergylett.7b00244

23. Y. Cao, A. Stavrinadis, T. Lasanta, D. So, G. Konstantatos, *Nat Energy*, 2016, **1**, 1–6, 10.1038/NENERGY.2016.35

24. A.G. Pattantyus-Abraham, I.J. Kramer, A.R. Barkhouse, X. Wang, G. Konstantatos et al., *ACS Nano*, 2010, **4**, 3374–80, 10.1021/nn100335g

25. J. Tang, K.W. Kemp, S. Hoogland, K.S. Jeong, H. Liu et al., *Nat Mater*, 2011, **10**, 765–71, 10.1038/nmat3118

26. A.H. Ip, S.M. Thon, S. Hoogland, O. Voznyy, D. Zhitomirsky et al., *Nat Nanotechnol*, 2012, **7**, 577–82, 10.1038/nnano.2012.127

27. K. Lu, Y. Wang, Z. Liu, L. Han, G. Shi et al., *Adv Mater*, 2018, **30**, 1–9, 10.1002/adma.201707572

28. D.A.R. Barkhouse, R. Debnath, I.J. Kramer, D. Zhitomirsky, A.G. Pattantyus-Abraham et al., *Adv Mater*, 2011, **23**, 3134–8, 10.1002/adma.201101065

29. W. Yoon, J.E. Boercker, M.P. Lumb, D. Placencia, E.E. Foos et al., *Sci Rep*, 2013, **3**, 1–7, 10.1038/srep02225

30. Y. Cao, A. Stavrinadis, T. Lasanta, D. So, G. Konstantatos, *Nat Energy*, 2016, **1**, 1–26, 10.1038/NENERGY.2016.35

31. N. Zhao, T.P. Osedach, L. yi Chang, S.M. Geyer, D. Wanger et al., *ACS Nano*, 2010, **4**, 3743–52,

32. J. Tang, L. Brzozowski, D.A.R. Barkhouse, X. Wang, R. Debnath et al., *ACS Nano*, 2010, **4**, 869–78, 10.1021/nn901564q

33. M. Sykora, A.Y. Koposov, J.A. Mcguire, R.K. Schulze, O. Tretiak et al., *ACS Nano*, 2010, **4**, 2021–34,

34. D. Bozyigit, S. Volk, O. Yarema, V. Wood, *Nano Lett*, 2013, **13**, 5284–8, 10.1021/nl402803h

35. G.W. Hwang, D. Kim, J.M. Cordero, M.W.B. Wilson, C.H.M. Chuang et al., *Adv Mater*, 2015, **27**, 4481–6, 10.1002/adma.201501156

36. A.R. Kirmani, A.D. Sheikh, M.R. Niazi, M.A. Haque, M. Liu et al., *Adv Mater*, 2018, **30**, 1–9, 10.1002/adma.201801661

37. J.M. Salazar-Rios, N. Sukharevska, M.J. Speirs, S. Jung, D. Dirin et al., *Adv Mater Interfaces*, 2018, **1801155**, 10.1002/admi.201801155

38. A.A. Bakulin, S. Neutzner, H.J. Bakker, L. Ottaviani, D. Barakel et al., *ACS Nano*, 2013, **7**, 8771–9, 10.1021/nn403190s

39. Z. Liu, S. Niu, N. Wang, *J Colloid Interface Sci*, 2018, **509**, 171–7, 10.1016/j.jcis.2017.09.010

40. R. Ihly, J. Tolentino, Y. Liu, M. Gibbs, M. Law, *ACS Nano*, 2011, **5**, 8175–86, 10.1021/nn2033117

41. S.R. Tulsani, A.K. Rath, *J Colloid Interface Sci*, 2018, **522**, 120–5, 10.1016/j.jcis.2018.03.047

42. H. Beygi, S.A. Sajjadi, A. Babakhani, J.F. Young, F.C.J.M. van Veggel, *Appl Surf Sci*, 2018, **457**, 1–10, 10.1016/j.apsusc.2018.06.152

43. W. Gao, G. Zhai, C. Zhang, Z. Shao, L. Zheng et al., *RSC Adv*, 2018, **8**, 15149–57, 10.1039/c8ra01422a

**Figures**

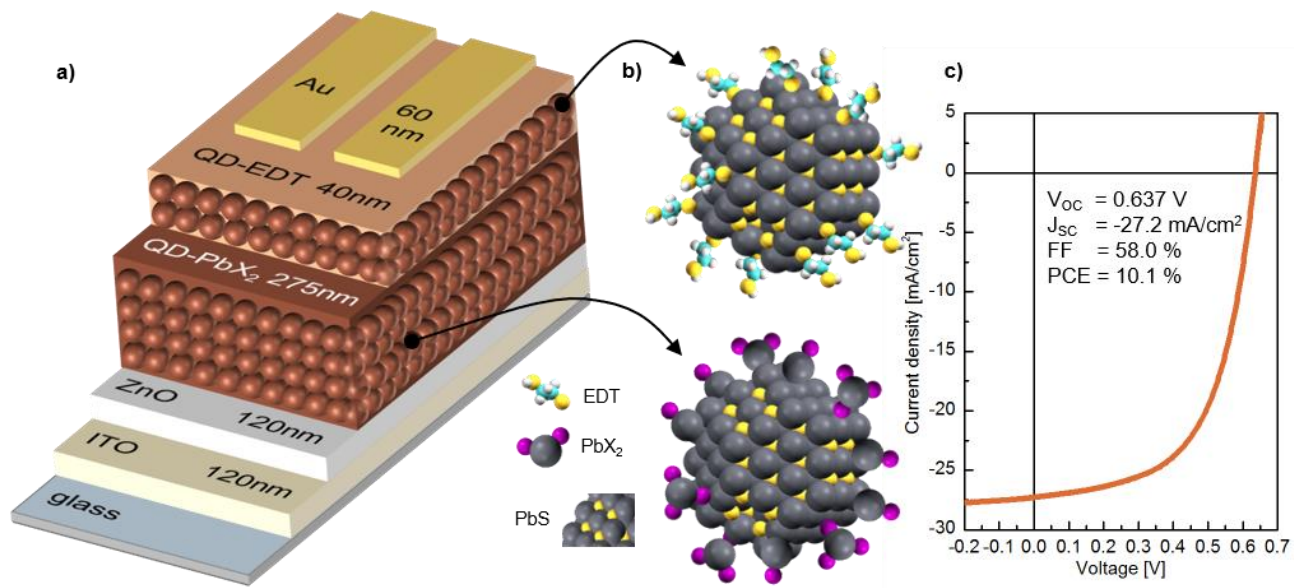

**Figure 1.** PbX$_2$-PbS QD solar cell structure and performance. **a)** Architecture of the layers in the thin-film devices, with thicknesses included. **b)** Quantum dots employed in the active and electron blocking layer. The PbS crystal structure is visible and the different ligand structures are depicted. **c)** I-V-characteristics of a, at maximum performance working PbS-PbX$_2$ QD device, while being exposed to oxygen.

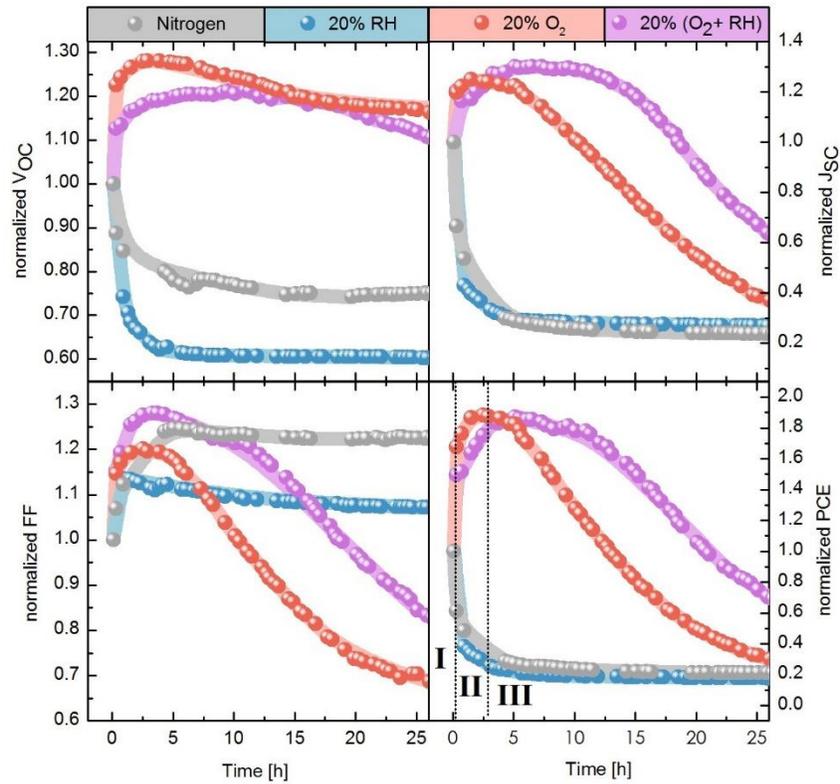

**Figure 2.** Evolution of photovoltaic parameters under different atmospheres. From left to right, the panels show the normalized Voc, Jsc, FF and PCE. The lines behind the data points are meant as a guide for the eye. In the PCE panel, the three phases of degradation under oxygen (I, II, and III) are defined. The first data point of each curve is always measured in nitrogen only. The gas flow is started after the first measurement.

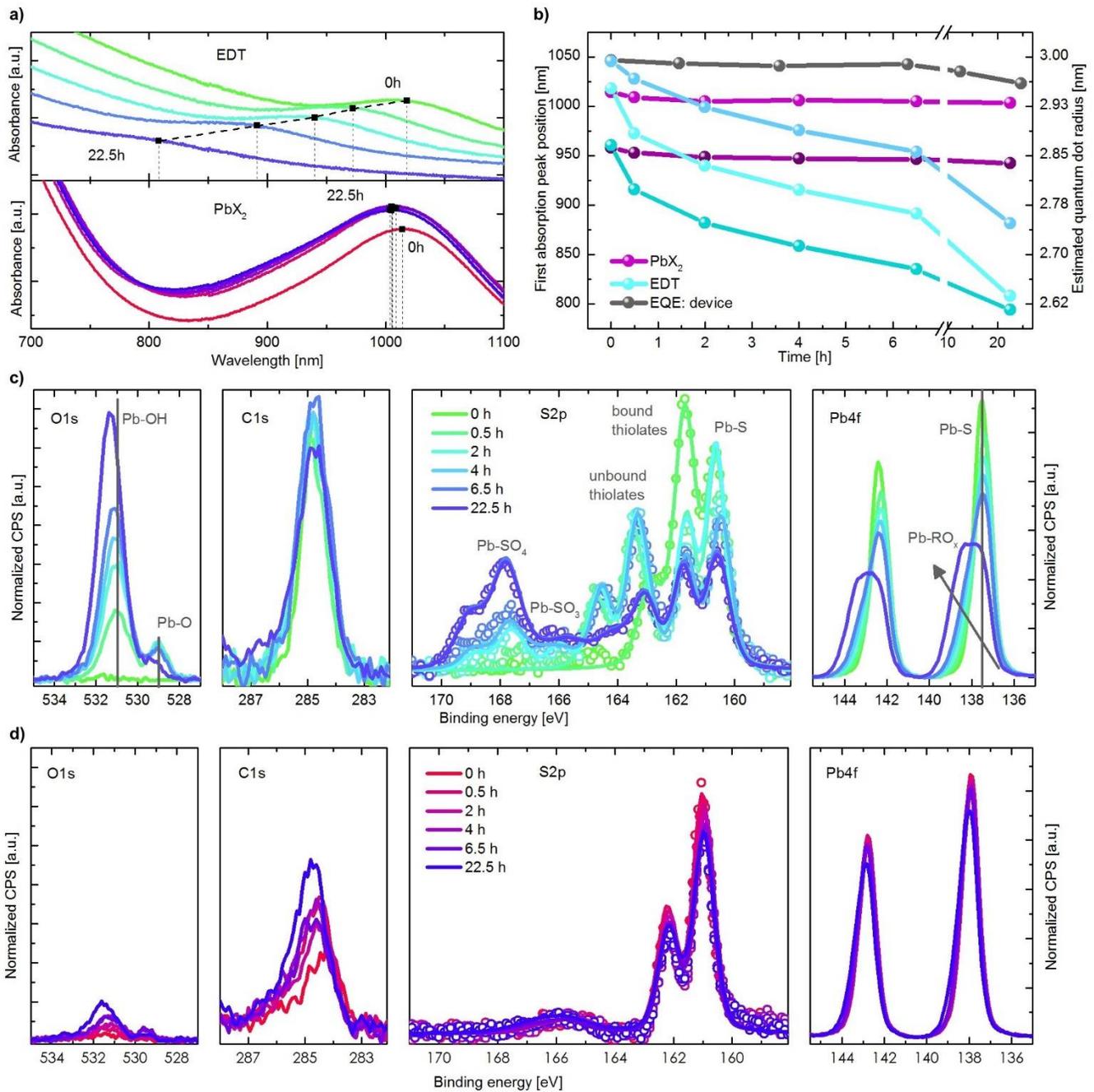

**Figure 3.** Initial QDs size dependence and ligand controlled degradation behaviour. Upper row from left to right: Absorption measurements of EDT and PbX$_2$ covered QDs for different time steps in an oxygen atmosphere. Extracted position of the first absorption peak maximum for different ligands and initial dot sizes, also including positions from an EQE measurement of a device. Middle row: XPS data of EDT-PbS

QDs for different times in an oxygen atmosphere. Lower row: XPS data in the same scale as the above graphs but for PbX$_2$-PbS QDs.

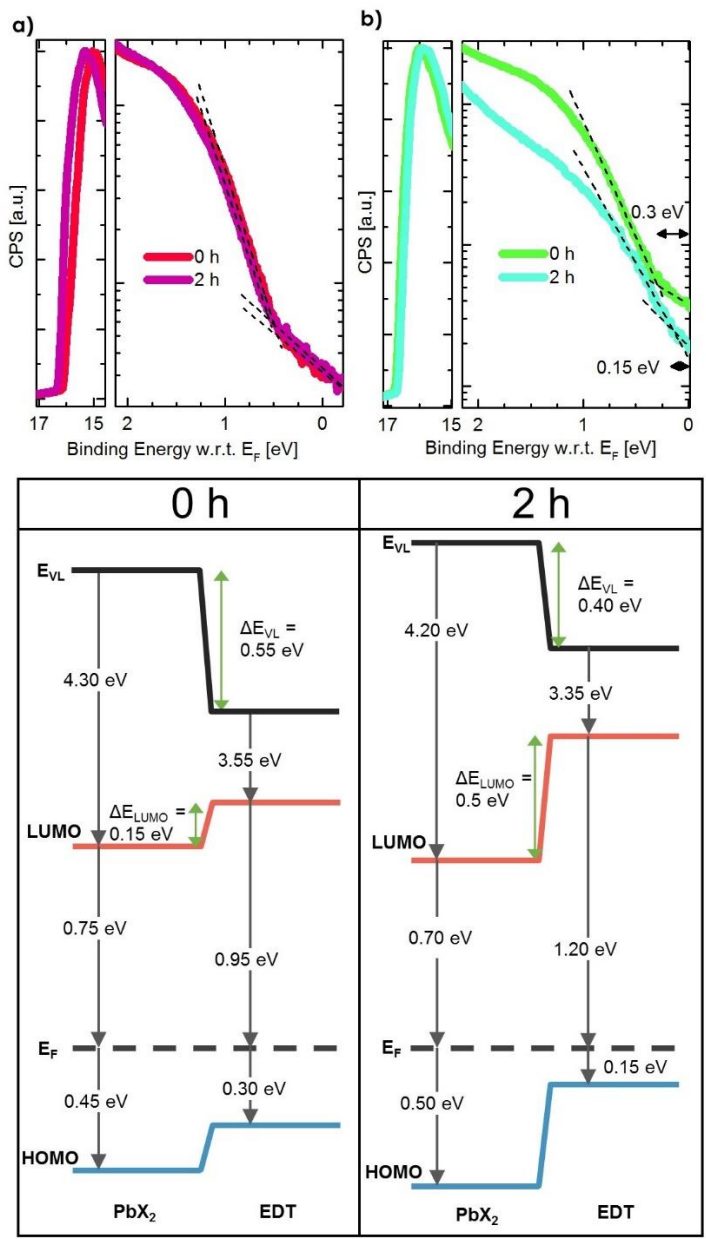

**Figure 4.** UPS data and energy diagrams for the degradation process. Upper row: Left, PbX$_2$-PbS QD film degraded for different times in oxygen. Right, the same measurement for EDT-PbS QDs. Lower row: For two time steps in an oxygen degradation run, the energetic of UPS and absorption data are combined to map the energetic landscape. The increased electron blocking barrier can easily be identified.

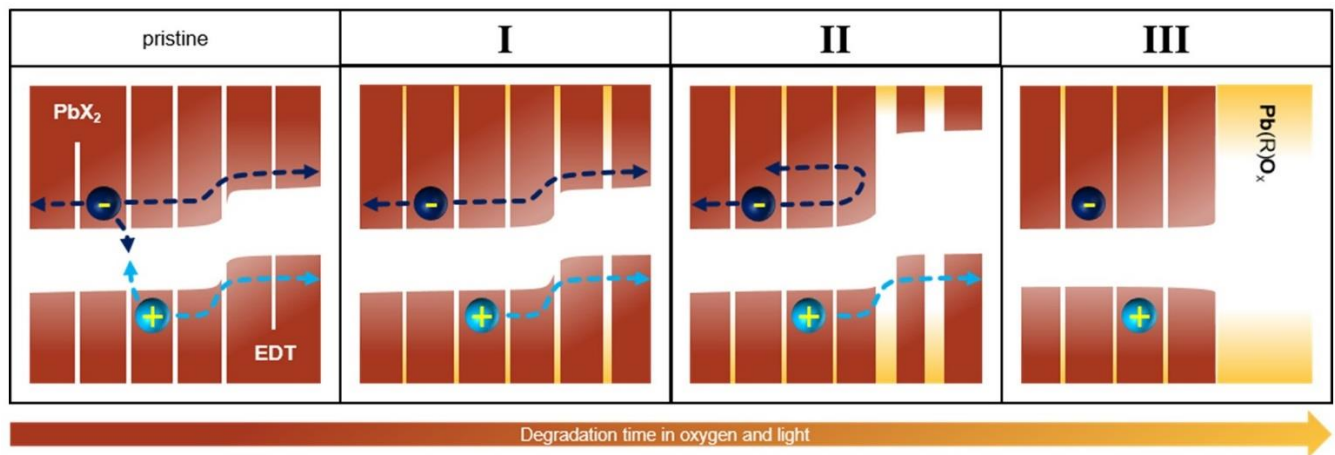

**Figure 5.** Energy landscape sketch to explain the increase in performance. The pristine and following phases are depicted from left to right. An exemplary electron and hole and their possible paths are shown. The HOMO and LUMO are depicted in red-brown for PbS and yellow for the degradation products. White space denotes the bandgaps.